\documentclass[a4paper,11pt]{article}
\usepackage{jheppub} 
\usepackage{lineno}
\newcommand{\bea}{\begin{eqnarray}}
\newcommand{\beq}{\begin{equation}}

\newcommand{\eea}{\end{eqnarray}}
\newcommand{\eeq}{\end{equation}}
\newcommand{\ds}{\displaystyle}

\usepackage{amsmath}
\usepackage{amsfonts}
\usepackage{amssymb}
\usepackage{bm}       
\usepackage[usenames,dvipsnames]{color}
\usepackage{comment}
\usepackage{dcolumn}  
\usepackage{epsfig}
\usepackage{graphics}
\usepackage{graphicx} 
\usepackage{lipsum}
\usepackage[rightcaption]{sidecap}

\usepackage{placeins} 
\usepackage{psfrag}
\usepackage{times}
\usepackage[varg]{txfonts}
\usepackage{xcolor}
\usepackage{xspace}

\graphicspath{%
    {converted_graphics/}
    {/}
}

\begin{document}

\title{{Generalised Non-Linear Electrodynamics: classical picture and 
 effective mass generation  }

}

\author[a,b,c]{Abedennour Dib }
\emailAdd{ abedennour.dib@cnrs-orleans.fr}
\affiliation[a]{Laboratoire de Physique et Chimie de l'Environnement et de l'Espace (LPC2E) UMR 7328, Centre National de la Recherche Scientifique (CNRS), Universit\'e d'Orl\'eans (UO) Centre National d'\'Etudes Spatiales (CNES), 3A Avenue de la Recherche Scientifique, 45071 Orl\'eans, France
}
\affiliation[b]{Observatoire des Sciences de l'Univers en region Centre (OSUC) UMS 3116
 Universit\'e d'Orl\'eans (UO), Centre National de la Recherche Scientifique (CNRS), Observatoire de Paris (OP), Universit\'e Paris Sciences \& Lettres (PSL), 1A rue de la F\'{e}rollerie, 45071 Orl\'{e}ans, France}
\affiliation[c]{D\'epartement de Physique, Unit\'e de Formation et Recherche Sciences et Techniques, Universit\'e d'Orl\'eans Rue de Chartres, 45100 Orl\'{e}ans, France}

\author[d]{Jos\'e A. Helay\"el-Neto }
\emailAdd{ helayel@cbpf.br}

\affiliation[d]{Departamento de Astrof\'{\i}sica, Cosmologia e Intera\c{c}\~{o}es Fundamentais (COSMO), Centro Brasileiro de Pesquisas F\'{\i}sicas (CBPF)
Rua Xavier Sigaud 150, 22290-180 Urca, Rio de Janeiro, RJ, Brazil}

\author[a,b,c]{Alessandro D.A.M. Spallicci  
}
\emailAdd{ spallicci@cnrs-orleans.fr}

\abstract{
Starting from a generic Lagrangian, we discuss the number of propagating degrees of freedom in the framework of generalised non-linear electrodynamics when a photon-background split is applied. We start by stating results obtained in a previous paper, before modifying the action to an equivalent form. Within this new formulation, we highlight the presence of an effective mass and consider the mechanical reduction of the model to ensure the positivity of said mass. We then study the constraint algebra of the model and show that we shift from a model with two first-class to two second-class constraints, which implies the propagation of an additional degree of freedom. We also show that the Hamiltonian is bound from below and thus does not suffer from Ostrogradski-type instabilities.  We conclude by deriving the propagator for the model, and discussing the potential link between the nature of this additional polarisation and the mechanism behind the effective mass generation in this class of models.}

\date{\today} 

\keywords{Vector Fields, Photons, Non-Linear Electrodynamics, Effective Mass, Hamiltonian Constraints, Classical Analysis}

\maketitle

\section{Introduction}
The inherently relativistic theory of electromagnetism is considered a pillar of modern physics. The adoption of the covariant language is usually considered the distinctive mark of Classical Electrodynamics (CED). 

The theoretical structure of CED expresses four features: the masslessness of the photon, the linearity of Maxwell's equations, the invariance under gauge and Lorentz-Poincar\'e transformations. The advent of Quantum Mechanics (QM) yielded the quantisation of electromagnetism, leading {  to} Quantum Electrodynamics (QED). QED is considered as one of the most precise models in theoretical physics, starting from the astonishing precision that predicted the value of the Lamb shift. Nonetheless, QED had to sacrifice one of the previously listed ingredients: the linearity of Maxwell's equations. Indeed it has been extensively shown that QED at a fundamental level is a non-linear theory, where the linear leading order requires non-linear corrections to obtain the proper results. This showed itself in terms of a limit that was first discovered by Sauter in 1931 \cite{Sauter-1931} and then further formalised in 1936 by Heisenberg and Euler \cite{Heisenberg-Euler-1936}. Today, the limit bears Schwinger's name for his calculation of the rate of pair production in 1951 \cite{Schwinger-1951}. 

A few years before Heisenberg and Euler's attempt at a non-linear model, Born and Infeld proposed their own non-linear extension to Maxwell's theory, that tried to fix the problem of the electron self energy at a classical level. The Born-Infeld model was also treated as a potential candidate for a unification between General Relativity and Electrodynamics, and has been extensively used in astrophysical and cosmological contexts \cite{Vollick-2003,Jafarzade-Zangeneh-Lobo-2021,Kim-2022,Cataldo-Garcia-1999,Cai-Pang-Wang-2004}. It also enjoyed a rebirth in the recent years, since it acts as a low energy limit of certain string theories \cite{Metsaev-Rahmanov-Tseytlin-1987}. 

Another well studied category of extensions of electrodynamics is that of  massive photons. Starting with de Broglie's and Proca's early work \cite{debroglie-1923,debroglie-1936,proca-1936d,proca-1937}, it has been an active area since. The idea of a photon mass is not necessarily in contradiction with experimental evidence as the third polarisation can be incredibly small and thus almost undetectable \cite{Ignatiev-Yoshi-1996}. On the other hand, it would simplify greatly the quantisation of electromagnetism and avoid the problem of infrared divergences due to soft photons \cite{Weinberg-1965}. 

While it is usually assumed to be zero, there can't be definitive proof of the masslessness of the photon thanks to Heisenberg's principle \cite{spallicci-benetti-capozziello-2022,Capozziello-Sarracino-Spallicci-2023}. Thus far, only upper bounds have been set for the value of its mass \cite{Navas-etal-2024,spallicci-etal-2024b}. It has been showed that in the Standard Model Extension (SME), a photon mass emerges \cite{bonetti-dossantosfilho-helayelneto-spallicci-2017,bonetti-dossantosfilho-helayelneto-spallicci-2018}, and models such as Bopp-Podolsky's \cite{bopp-1940} which include higher derivatives are shown to contain two modes, one massive and one massless. An effective photon mass is also used in the context of condensed matter, to explain certain phenomena such as the Meissner effect in superconductivity
\cite{Weinberg-1986} and in astrophysics \cite{helayelneto-spallicci-2019,spallicci-etal-2021}. 

Approaches that mix both extensions also exist in the literature. As of recently, there has been a certain quantity of papers that treat the Generalised Proca class of models \cite{Heisenberg-2014}, one of which is the recent Proca-nuevo \cite{derham-etal-2023} model. These extensions notably allow to cross the bridge with massive gravity \cite{derham-etal-2022,defelice-etal-2016}, which is another active area of research. The work presented in this paper is a part of this class of extensions.

The goal is to show that within the framework offered by Generalised Non-Linear Electrodynamics (GNLED) it is possible for a massive mode to emerge. 

\bibliographystyle{apsrev} 

We will start by introducing GNLED as a framework, with its postulates and main results. We will then show that it is possible to break the manifest U(1) symmetry at the action level. We will complete this description by studying the Hamiltonian structure of the model, its constraints algebra and their nature as well as the emergence of the mass and its expression. We conclude with a discussion and  perspectives.

In this paper, we work in Minkowski spacetime, where the metric has as a convention $\eta^{\mu\nu}=\left(+,-,-,-\right)$. Spacetime indices are labelled with Greek symbols, time indices as $0$ components and space indices with Latin symbols.
  
\section{Lagrangian prescription}
\subsection{Construction of the action}
GNLED is a class of electromagnetic models that are constructed as a power expansion of Maxwell invariants. It can be shown that these models are general in the sense that already known extensions of electrodynamics are already encompassed within the framework. This class of models has already been extensively discussed in \cite{spallicci-etal-2024a} and the following is thus a brief introduction. The two Maxwell invariants are defined as  
\begin{align}
   & \mathcal{F}=-\frac{1}{4\mu_0}F^{\mu\nu}F_{\mu\nu}=\frac{1}{2\mu_0}\left(\frac{\Vec{E}^2}{c^2}-\Vec{B}^2\right)~,  &
   \\
  & \mathcal{G}=-\frac{1}{4\mu_0}F^{\mu\nu}\Tilde{F}_{\mu\nu}=-\frac{1}{4\mu_0}F^{\mu\nu}\epsilon_{\mu\nu\rho\sigma}F^{\rho\sigma}=\frac{1}{\mu_0 c}\Vec{E}\cdot\Vec{B} ~,&
\end{align}
where $\Vec{E}$ is the electric field, $\Vec{B}$ the magnetic field and $F^{\mu\nu}$ the field strength tensor.

Let us now consider the Born-Infeld and the Heisenberg-Euler actions. In the weak-field limit \cite{Sorokin-2022}, they can be expressed as 
\begin{align}
&\mathcal{L}_{BI}=-\frac{1}{4\mu_0} F^{\mu\nu}F_{\mu\nu}+\frac{1}{32b}\left[\left(F^{\mu\nu}F_{\mu\nu}\right)^2+ \left(\Tilde{F}^{\mu\nu}F_{\mu\nu}\right)^2\right]~, &
\\
&\mathcal{L}_{EH}=-\frac{1}{4\mu_0} F^{\mu\nu}F_{\mu\nu} +\beta\left[\left(F^{\mu\nu}F_{\mu\nu}\right)^2+ 7\left(\Tilde{F}^{\mu\nu}F_{\mu\nu}\right)^2\right]~,& 
&
\end{align}
where $b$ and $\beta$ are constants. It can be shown that additional orders allow to deal with the strong field limit.
From this, we generalise the procedure and build a Lagrangian that can then be expressed in terms of integer powers of the two invariants
\begin{align}
        \mathcal{L}(\mathcal{F},\mathcal{G})=&\mathcal{L}_0+\frac{\partial \mathcal{L}}{\partial \mathcal{F}}\mathcal{F}+\frac{\partial \mathcal{L}}{\partial \mathcal{G}}\mathcal{G}+ \frac12 \frac{\partial^2 \mathcal{L}}{\partial {\mathcal{F}}^2}\mathcal{F}^2+\frac12 \frac{\partial^2 \mathcal{L}}{\partial {\mathcal{G}}^2}\mathcal{G}^2+ \frac{\partial^2 \mathcal{L}}{\partial \mathcal{F}\partial \mathcal{G}}\mathcal{F}\mathcal{G}+...&~.
\end{align}

This family of models in its initial formulation falls under the Plebanski class of electrodynamics: models that are both Lorentz invariant and gauge invariant under U(1) transformations \cite{Plebankski-1970, Schellestede-Perlick-Lammerzahl-2016}. As such, they do not admit massive extensions. A way to exit the Plebanski class is to split the electromagnetic field into a strong background and a photon field that acts as a perturbation, {  indicating them with} { upper and lower case, respectively. With this new notation, we} 
start by defining the field strength tensor as
\begin{align}
    F^{\mu\nu}_{Total}={F}^{\mu\nu}+f^{\mu\nu} ~,
\end{align}
where { $F^{\mu\nu}=\partial_\mu A_\nu-\partial_\nu A_\mu$ and $f^{\mu\nu}=\partial_\mu a_\nu-\partial_\nu a_\mu$ are respectively the background and photon field-strength tensor. For the Maxwell invariants, it gives} 
\begin{align}
   & \mathcal{F}_{Total}= {F}^{\mu\nu}{F}_{\mu\nu}+2 {F}^{\mu\nu}f_{\mu\nu}+ f^{\mu\nu}f_{\mu\nu}= \mathcal{F}+\delta \mathcal{F}_{Total} ~,& \\
   & \mathcal{G}_{Total}= \tilde{F}^{\mu\nu}{F}_{\mu\nu}+2 \tilde{F}^{\mu\nu}f_{\mu\nu}+\tilde{f}^{\mu\nu}f_{\mu\nu}= \mathcal{G}+\delta \mathcal{G}_{Total}~. & 
\end{align}

 It is assumed that said expansion is well defined, {\it{i.e.}} that it converges and that each order contributes less than the previous one. It is also required for the background field to be non-dynamical. Before going further, we will shift into the natural unit system where $h=c=\mu_0=1$. The resulting Lagrangian at the second order would then be 
\begin{align}
    &\mathcal{L}(\mathcal{F},\mathcal{G})_{Total}=\mathcal{L}(\mathcal{F}+\delta \mathcal{F}, \mathcal{G}+\delta \mathcal{G})=\mathcal{L}(\mathcal{F},\mathcal{G})+\sum_{n=1}^i\frac{1}{n!}\left(\delta \mathcal{F}\frac{\partial}{\partial\mathcal{F}}+\delta \mathcal{G}\frac{\partial}{\partial \mathcal{G}}\right)^n\mathcal{L} \Big|_B \nonumber \\
    & \approx \mathcal{L}(\mathcal{F},\mathcal{G})+\delta \mathcal{F}\left.\frac{\partial\mathcal{L}}{\partial \mathcal{F}}\right\vert_B+\delta \mathcal{G}\left.\frac{\partial\mathcal{L}}{\partial \mathcal{G}}\right\vert_B  +\frac12\left(\delta \mathcal{F}\right)^2\left.\frac{\partial^2\mathcal{L}}{\partial \mathcal{F}^2} \right\vert_B
    +\frac12\left(\delta \mathcal{G}\right)\left.\frac{\partial^2\mathcal{L}}{\partial \mathcal{G}^2}\right\vert_B + \delta \mathcal{F}\delta \mathcal{G}\left.\frac{\partial^2\mathcal{L}}{\partial \mathcal{F}\partial \mathcal{G}}\right\vert_B=\mathcal{L}_{B}+ \mathcal{L}_{2} ~, &
\end{align}
where $\mathcal{L}_{B}$ represents the entirety of the pure background contributions and  $\mathcal{L}_2$ contains the photon terms and photon-background interactions. We also define the coefficients 
\begin{align}
    &C_1=\left.\frac{\partial\mathcal{L}}{\partial \mathcal{F}}\right\vert_B~, &C_2=\left.\frac{\partial\mathcal{L}}{\partial \mathcal{G}}\right\vert_B ~,&&D_1=\left.\frac{\partial^2\mathcal{L}}{\partial \mathcal{F}^2}\right\vert_B \nonumber ~,&\\ 
    &\left.D_3=\frac{\partial^2\mathcal{L}}{\partial \mathcal{G}^2}\right\vert_B ~,&\left.D_2=\frac{\partial^2\mathcal{L}}{\partial \mathcal{F}\partial \mathcal{G}}\right\vert_B~,&\nonumber
\end{align}
which gives for $\mathcal{L}_2$
\begin{align}\label{actionorder2}
    \mathcal{L}_2=&-\frac12\left(C_1{F}^{\mu\nu}+C_2{\tilde{F}}^{\mu\nu}\right)f_{\mu\nu}-\frac14 C_1f^{\mu\nu}f_{\mu\nu}-\frac14 C_2 f^{\mu\nu}\tilde{f}_{\mu\nu}+\frac18 \left({K}^{\mu\nu\rho\sigma}+2{T}^{\mu\nu\rho\sigma}\right)f_{\mu\nu}f_{\rho\sigma}~,
\end{align}
where $K$ and $T$ are the shorthand notation for 
\begin{align}
    &{K}^{[\mu\nu][\rho\sigma]}=D_1{F}^{\mu\nu}{F}^{\rho\sigma}+D_3 {\tilde{F}}^{\mu\nu}{\tilde{F}}^{\rho\sigma}~~,&\\
    &{T}^{[\mu\nu][\rho\sigma]}=D_2{F}^{\mu\nu}{\tilde{F}}^{\rho\sigma}~~,& 
\end{align}
the indices $\{\mu,\nu\}$ and $\{\rho,\sigma\} $ are antisymmetric within their respective brackets, $\left[\mu\nu, \rho\sigma\right]$ are symmetric between themselves; finally,  $\{\mu,\rho\}, \{\mu,\sigma\}, \{\nu,\rho\},\{\nu,\sigma\}$ do not commute.

The variation of the action $\delta \mathcal{L}_2 $ 
\begin{align}
    \delta \mathcal{L}_2= &-\left(C_1{F}^{\mu\nu}+C_2{\tilde{F}}^{\mu\nu}\right) \partial_\mu \delta a_\nu-\left(C_1 f^{\mu\nu}+C_2 \tilde{f}^{\mu\nu}\right)\partial_\mu \delta a_\nu +\frac12\left({K}^{\mu\nu\rho\sigma}+2{T}^{\mu\nu\rho\sigma}\right)f_{\rho\sigma} \partial_\mu \delta a_\nu ~~,& 
\end{align}
with respect to the photon 4-potential $a_\mu$, gives the field equations 
\begin{align}
     &-\partial_\mu\left(C_1 {F}^{\mu\nu}+C_2{\tilde{F}}^{\mu\nu}\right)-\partial_\mu \left(C_1 f^{\mu\nu}+C_2 \tilde{f}^{\mu\nu}\right)+\frac12\partial_\mu\left({K}^{\mu\nu\rho\sigma}f_{\rho\sigma}+2{T}^{\mu\nu\rho\sigma}f_{\rho\sigma}\right)=0~,&
\end{align}
which constitute the modified Maxwell equations \cite{spallicci-etal-2024a}.
We now define $S^{\mu\nu\rho\sigma}$ and $R^{\mu\nu}$ as the following
\begin{align}
   & S^{\mu\nu\rho\sigma}={K}^{\mu\nu\rho\sigma}+2{T}^{\mu\nu\rho\sigma}~,&\\
   &R^{\mu\nu}=C_1{F}^{\mu\nu}+C_2{\Tilde{F}}^{\mu\nu}~.&
\end{align}
\subsection{Field redefinition}
  Looking back at the action described by \eqref{actionorder2}, we can see that there exist no free Maxwellian term even at first order in Maxwell invariants,  represented by the terms coupled to $C_1$ and $C_2$.
We shall discuss this further by looking solely at the first order action
\begin{align}
    \mathcal{L}_1= -\frac12\left(C_1{F}^{\mu\nu}+C_2{\tilde{F}}^{\mu\nu}\right)f_{\mu\nu}-\frac14 C_1f^{\mu\nu}f_{\mu\nu}
    &-\frac14 C_2 f^{\mu\nu}\tilde{f}_{\mu\nu}~.
\end{align}

For the sake of simplicity we shall drop {  the} dual contributions by setting {$C_2=0$}, {  and thus} yielding
\begin{align} \label{L1}
       \mathcal{L}_1= -\frac14 C_1 f^{\mu\nu}f_{\mu\nu} - \frac12 C_1 F^{\mu\nu}f_{\mu\nu}~,
\end{align}
 which can be brought into the following form
\begin{align}\label{ActionOrder1}
       \mathcal{L}_1= -\frac14 C_1 f^{\mu\nu}f_{\mu\nu} + a_\nu \mathcal{J}^\nu~,
\end{align}
where $\mathcal{J}^\nu= \partial_\mu \left(C_1 F^{\mu\nu}\right)$  assumes the role of a current. Looking at the equations of motion for \eqref{L1}
\begin{align}
    \partial_\mu (C_1f^{\mu\nu})+\mathcal{J^\nu}=0~,
\end{align}
from the first term, we {  infer} 
that in the case of a non constant background, the $C_1$ coefficient has a non trivial effect on the photon propagation, and in a general case does not lead to a free propagating photon, as it is fundamentally coupled to the background.\\

For the following we will only consider
\begin{align}
    \mathcal{L}_0=- \frac14 C_1 f^{\mu\nu}f_{\mu\nu}~~,
\end{align}
to ensure a trivially free photon at first order, we could 
apply the following field redefinition
\begin{align}
    f'^{\mu\nu}\rightarrow \sqrt{C_1} f^{\mu\nu}~, \nonumber
\end{align}
which would lead to an action of the form
\begin{align}
    \mathcal{L}'_0= -\frac14 f'^{\mu\nu}f'_{\mu\nu}~,\nonumber
\end{align}
giving us the desired free propagating photon. However, this field redefinition would be ill advised. Indeed, the 2-form $f^{\mu\nu}$ is not the field dictating the dynamics of the model. This role is imputed to the 4-potential $a^\mu$, and as such, a field redefinition must act on the potentials. The $C_1$ coefficient not being a constant, but a function of the background field, 
{  will} necessarily lead to radically different results.

Let us thus consider the following redefinition
\begin{align} \label{fieldredif}
    a'^\mu \rightarrow \alpha ~a^\mu~, 
\end{align}
{  with} 
$\alpha^2= C_1$. Constructing a field strength tensor for the redefined potential $a'^\mu$ gives
\begin{align}
f'^{\mu\nu}=\partial^\mu a'^\nu - \partial^\nu a'^\nu = \alpha f^{\mu\nu}+ a^\nu \partial^\mu \alpha- a^\mu \partial^\nu \alpha&~~,
\end{align}
and thus
\begin{align}
    \alpha f^{\mu\nu}=f'^{\mu\nu}-\frac{1}{\alpha}\left( a'^\nu \partial^\mu \alpha- a'^\mu \partial^\nu \alpha\right)~~,
\end{align}
defining $V^\mu = {\ds \frac{\partial^\mu \alpha}{\alpha}}$, we get the following expression
\begin{align}
    \alpha f^{\mu\nu}=f'^{\mu\nu}-\left( a'^\nu V^\mu- a'^\mu V^\nu\right)&~~.
\end{align}

It is very easy to see that this field redefinition breaks the gauge invariance of the $a'^\mu$ field if $\alpha$ is not a constant. While field redefinitions that break gauge symmetry were considered a no-go, recent work \cite{Cohen-Forslund-Helset-2024} demonstrated that the space of allowed field redefinitions is much larger than initially thought, and that symmetry breaking redefinitions are acceptable.  
 We could consider that the effects of the background at the level of the potentials rather than the field-strengths, induce a symmetry breaking. We shall provide more details on this argument in the discussion section. 

From this discussion, we 
derive 
\begin{align}
     \alpha^2 f^{\mu\nu}f_{\mu\nu}=f'^{\mu\nu}f'_{\mu\nu}+ 4 a'_\mu V_\nu f'^{\mu\nu}+2 \left(a'^\nu a'_\nu V^\mu V_\mu- a'_\mu a'_\nu V^\mu V^\nu\right)&~~,
\end{align}
and thus get the equivalent form of the action
\begin{align}\label{action 1}
    \mathcal{L}'_0= -\frac14f'^{\mu\nu}f'_{\mu\nu}-a'_\mu V_\nu f'^{\mu\nu}-\frac12 \left(a'^\nu a'_\nu V^\mu V_\mu- a'_\mu a'_\nu V^\mu V^\nu\right)~~.
\end{align}

Going back to the action described by \eqref{ActionOrder1}, we can show that the second term will still behave as a current term {  post-field redefinition}
\begin{align}
     a_\nu\mathcal{J}^\nu \rightarrow \frac{1}{\alpha} a'_\nu \mathcal{J}^\nu = a'_\nu \mathcal{J}'^\nu ~,
\end{align}
with $\mathcal{J}'^\nu ={\ds \frac{\partial_\mu \left(\alpha^2 F^{\mu\nu}\right)}{\alpha}}$.
\subsection{Equations of motion and gauge invariance}

In the class of models we are discussing, the background field is non-dynamical, and as such $V^\mu$ has to be a purely space-like vector since $\partial^0 \alpha=0$. Additionally, our following results will be derived in a context where we chose $V^\mu= cst$. This can be justified in different manners, either by choosing a specific type of background that trivially ensures this property, or by demanding that at the first order in its Taylor expansion the $V^\mu$ vector is constant .

Computing the variation of the action with respect to the four potential $a'\mu$ yields the equations of motion
\begin{align} \label{EOM'1}
    \partial_\mu f'^{\mu\nu}+ V^\nu \partial^\mu a'_\mu - V^\mu \partial_\mu a'^\nu-a'^{\nu} V^\mu V_\mu+ a'^\mu V^{\nu}V_\mu=0~~,
\end{align}
the divergence of \eqref{EOM'1} yields the following subsidiary condition
\begin{align}
    V_\nu \partial_\mu f'^{\mu\nu}+ V^2 \partial_\mu a'^\mu -V^\mu V^\nu \partial_\mu a'_\nu=0~~,
\end{align}
notice that this subsidiary condition can also be obtained by multiplying the EOM by $V_\nu$, as such they are not fully independent.

Having a closer look at the gauge invariance of the action and using the usual gauge transformation for $U(1)$ theories $a''^\mu= a'^\mu + \partial^\mu \chi$  we get
\begin{align}
    &\delta \mathcal{L}= \delta\left(-\frac14f'^{\mu\nu}f'_{\mu\nu}-a'_\mu V_\nu f'^{\mu\nu}-\frac12 \left(a'^\nu a'_\nu V^\mu V_\mu- a'_\mu a'_\nu V^\mu V^\nu\right)\right) \nonumber\\
   & \delta \mathcal{L}= \left(  V_\nu \partial_\mu f'^{\mu\nu}+ V^2 \partial_\mu a'^\mu -V^\mu V^\nu \partial_\mu a'_\nu\right) \chi~~,
\end{align}
which holds up to 4-divergences. This result confirms that the action is not gauge invariant. Indeed, even though the subsidiary condition is not independent from the equations of motion, gauge invariance has to upheld off-shell.

However, while the action is not gauge invariant, it is always possible to add a Stueckelberg field \cite{ruegg-ruizaltaba2004} such as $  a'^\mu \rightarrow a'\mu+ \partial^\mu \phi $ and retrieve the gauge invariance under the following transformations
\begin{align}
    &\delta a'^\mu= \partial^\mu \Lambda~~,\\ 
    &\delta \phi =-\Lambda ~~.
\end{align}
\subsection{Mechanical reduction}
The term $ V^\mu V^\nu$ is a totally symmetric tensor, and can be decomposed in its fundamental representation as
\begin{align}
    V^\mu V^\nu= v^\mu v^\nu-\frac14 \eta^{\mu\nu} V_\alpha V^\alpha~~,
\end{align}
where $v^\mu v^\nu$ is a totally symmetric and traceless tensor. Again for simplicity, we will set this tensor to be zero. It is however very easy to see that the results would hold despite this condition in a more general scenario as this will be demonstrated in the last section of this paper.  
From this we get in the action 
\begin{align}
     \mathcal{L}_0= -\frac14f'^{\mu\nu}f'_{\mu\nu}-a'_\mu V_\nu f'^{\mu\nu}-\frac58 V^{\mu}V_\mu a'^\nu a'_\nu~~,
\end{align}
since $V^\mu$ is purely spacelike, we have  $V^\mu V_\mu= V^0V_0+V^iV_i= -V_iV_i=-\vec{V}^2$ which in turn allows us to define $ M = \frac54 \vec{V}^2$  and get
\begin{align}\label{ActionMass}
    \mathcal{L}_0= -\frac14f'^{\mu\nu}f'_{\mu\nu}-a'_\mu V_\nu f'^{\mu\nu}+ \frac12 M a'^\nu a'_\nu~~,
\end{align}
ensuring the proper sign for the mass term. 
From this, we can get the following form for the equations of motion 
\begin{align}
    \partial_\mu f'^{\mu\nu}+ M a'^\nu+ V^\mu {f'_\mu}^\nu +V^\nu \partial_\mu a'^\mu=0~~,
\end{align}
which is equivalent to \eqref{EOM'1} up to a 4-divergence.

Expliciting the potentials
\begin{align}
    \left(\eta^{\mu\nu}\Box - \partial^\mu \partial^\nu+ \eta^{\mu\nu} M+\eta^{\mu\nu} V_\alpha \partial^\alpha-V^\mu \partial^\nu+ V^\nu \partial^\mu \right) a'_\mu=0 ~~,
\end{align}
and using the phase exponential ansatz $ \partial^\mu = -ik^\mu$
\begin{align}
    &\left(-\eta^{\mu\nu}k^2+ k^\mu k^\nu+ \eta^{\mu\nu} M-i\eta^{\mu\nu} V_\alpha k^\alpha+iV^\mu k^\nu- V^\nu k^\mu \right) \tilde{a}'_\mu=0 ~~,
\end{align}
\begin{align}
    &D^{\mu\nu}~\tilde{a}'_\mu=0~~.
\end{align}

To find the dispersion relations, we have to compute the determinant of the $D^{\mu\nu}$ matrix. A simpler case is considering the mechanical reduction of the model, where $k^i=0$. 

Keeping in mind that $V^0=0$ we can expand the equations of motion into 
\begin{align}
    &-\tilde{a}'_0 M-i V^ik^0\tilde{a}'_i=0~,\label{Mechred1}\\ \nonumber\\
    &-k^2\tilde{a}'_i+ M\tilde{a}'_i- iV_ik^0\tilde{a}'_0=0~, \label{Mechred2}
\end{align}

From \eqref{Mechred2}
\begin{align}
    \tilde{a}'_i = \frac{iV^ik^0}{-k^2+M}\tilde{a}'_0 ~~,\label{Mechred3}
\end{align}
and \eqref{Mechred1}
\begin{align}
    \tilde{a}'_0= -\frac{iV^ik^0}{M}\tilde{a}'_i ~~,\label{Mechred4}
\end{align}
injecting \eqref{Mechred3} into \eqref{Mechred1} and \eqref{Mechred4} into \eqref{Mechred2}, it gives, respectively,
\begin{align}
    &\left(\frac95 k_0^2M-M^2\right)\tilde{a}'_0=0 ~~,\\ \nonumber\\
   &\left(-\frac95 k_0^2 + M\right) \tilde{a}'_i=0~~,
\end{align}
and as such we get for the frequency 
\begin{align}
    k_0=\omega = \sqrt{\frac59M}~~,
\end{align}
which appears to be definite positive due to the definition of $M$ ensuring the absence of tachyonic modes at the classical level.

From \eqref{ActionMass} it is also possible to apply the Hilbert procedure and derive the symmetric energy momentum tensor of the model which gives
\begin{align}
    T_{\mu\nu}=&{f'_\mu} ^{\alpha}{f'_\nu}_{\alpha}+\left(a'_\mu V_\alpha-a'_\alpha V_\mu\right){f'_\nu} ^{\alpha}+\left(a'_\nu V_\alpha-a'_\alpha V_\nu\right){f'_\mu} ^{\alpha} -a'_\mu a'_\nu M + \eta_{\mu\nu}\mathcal{L}_0 ~~.
\end{align}

The mechanical reduction is somewhat equivalent to the rest frame of a system, and seems to indicate the presence of an effective mass term. This is however, insufficient to conclude on its existence, and the count of propagating degrees of freedom has not been made. This could indeed simply be a screening effect, and to conclude on this idea we will now shift to the Hamiltonian picture to analyse the constraint algebra and count the physical degrees of freedom.

\section{Hamiltonian prescription and degrees of freedom}
We start by rewriting the Lagrangian in a fashion that breaks the manifest covariance. 
{  {\begin{align}
    \mathcal{L}_0= &-\frac12 f'^{0i}f'_{0i}-\frac14 f'{ij}f'_{ij}+ \frac12M \left(a'^0 a'_0- a'^i a'_i\right)-a'_0V_i f'^{0i}+a'_i V_j f'^{ij}
\end{align}}}

The next step is to define the phase space variables. The background fields being non-dynamical implies that the phase space has the same parameters as the classical EM one, {  \it i.e.} the potential $a'^{\mu}$ and its conjugate momentum $\pi^{\mu}$.  

  \begin{align}
  & \pi^\mu =\frac{\partial \mathcal{L_0}}{\partial \dot{a'_\mu}}=\! \begin{cases}
       & \!\!\!\!\! \pi^0= \frac{\ds\partial \mathcal{L}_0 }{\ds\partial \partial_0 a'_0}=0\\\\
      & \!\!\!\!\! \pi^i= \frac{\ds\partial \mathcal{L}_0}{\ds\partial \partial_0 a'_i}=f'^{i0}-a'^0V^i \nonumber
    \end{cases}& ~~,
\end{align}
where $\pi^0=0$ is a primary constraint that indicates the absence of propagating time-like polarisation. After the Legendre transform, we get 
{{  {\begin{align}
    \mathcal{H}=& \int d^3 x ~\left[\pi^i\partial_0 a'_i- \mathcal{L}_0\right]&\nonumber\\ 
    =&\int d^3x ~ \left[ \pi^i\partial_i a'_0 - \pi^iV_ia'_0-\pi^i \pi_i- \mathcal{L} _0 \right]\nonumber\\
    =& \int d^3x ~ \left[- a'_0\partial^i \pi_i- \pi^iV_ia'_0-\frac12 \pi^i\pi_i+ \frac14 f'^{ij}f'_{ij}+a'_i V_j f'^{ij}- \frac12\left(M'a'^0a'_0+ M a'^i a'_i \right)\right]~~,
\end{align}
where $M' = \frac94 V^iV_i$.
}}}

With the Hamiltonian density in hands, we can now study the constraints. For this, we will apply the Dirac-Bergmann algorithm \cite{Dirac-1950,Anderson-Bergmann-1951} to move from a model with singular Legendre transform into a fully constrained Hamiltonian system \cite{Bergmann-Goldberg-1955}. We will check the constraints consistency and then add all additional constraints into an augmented Hamiltonian with Lagrange multipliers. Said Hamiltonian will be of the form
{  {\begin{align}
    \mathcal{H}_{Aug}=\int d^3 x \mathcal{H} + \lambda_i \phi^{i}~,
\end{align}}}
where $\phi^i$  represents all the constraints and $\lambda_i$ are the Lagrange multipliers that enforces them.
{{  {We have $\pi^0=0$ acting as a constraint which we will define as
\begin{align}
    \phi_1=\pi^0 \approx 0 ~~,
\end{align}}}}
and as such

  {\begin{align}
    \mathcal{H}_{Aug}=\int d^3 x \mathcal{H} + \lambda^1 \phi_1~ .
\end{align}}

The first step is to evaluate the Poisson Bracket of the primary constraint $\phi_1$ with the augmented Hamiltonian
\begin{align}
       \{ \phi_1~,~\mathcal{H}_{aug}\}=\{ \pi^0~,~\mathcal{H}_{aug}\}= \partial^i\pi_i +\pi^iV_i+a'_0 M' \approx0~~,
\end{align}
where the notion of weak equality \cite{Wipf-2005} is used to ensure the consistency of the constraint. The result obtained above is a new constraint, which is nothing more than the modified Gauss law for the system. It will be referred to as
{  {\begin{align}
    \phi_2=\partial^i\pi_i +\pi^iV_i+a'_0 M' \approx0~~,
\end{align}}}
for the rest of the paper.

We now need to add this constraint to the augmented Hamiltonian before continuing further, which yields 
{  {
\begin{align}
    \mathcal{H}'_{aug}= \mathcal{H}+ \int d^3x ~ \lambda^1\phi_1+ \lambda^2 \phi_2~~,
\end{align}}}
computing the Poisson Bracket (PB) between the new constraint and $\mathcal{H}'_{aug}$ gives
{  {\begin{align}
    \{ \phi_2~,~\mathcal{H}'_{aug}\}=&\frac32 V_k\partial_jf'^{jk}+V_kV^k\partial_j a'^j-V^iV^k\partial_k a'_i-M\partial_j a'^j+ \lambda_1 M' \approx0 \nonumber\\
    &\Rightarrow \lambda_1=-\frac{5}{6 M}\partial_j f'^{jk}- \frac{1}{27} \partial^j a'_j~~,
\end{align}}}
which only fixes a Lagrange multiplier. There are thus no additional constraints. 
\subsection{Constraints algebra}
As we now possess the full set of constraints, we can derive their algebra following Dirac's classification \cite{Dirac-1950}
\begin{itemize}
    \item Constraints which have their Poisson Bracket vanish with all other constraints are referred to as First Class. They are known in the literature as the generators for gauge transformations \cite{Henneaux-Teitelboum-1992}.
    \item Constraints which have their Poisson Bracket with at least one other constraint not vanish are referred to as Second Class. They are described as physical degrees of freedom rather than redundancies in the theory, or as gauge fixations \cite{Henneaux-Teitelboum-1992}.
\end{itemize} 

The procedure is as described very straightforward and gives 
{  {\begin{align}
    \{\phi_1 (t,\vec{x})~,~\phi_2(t,\vec{y})\}=-M' \delta^3(\vec{x}-\vec{y}) \neq0~~,
\end{align}}}
which indicates that the system is composed of two second class constraints that form a skew-symmetric invertible matrix
\begin{align}
  \phi^{ij}(\vec{x},\vec{y}) = \left(\begin{array}{cc}
         0&  -M'\\
         M'& 0
    \end{array}\right) \delta^3(\vec{x}-\vec{y})~~,
\end{align}
obeying the identity 
\begin{align} \label{constraintID}
    \int d^3 z \sum_{k} \phi_{ik}(\vec{x},\vec{z})~\phi^{-1}_{kj}(\vec{z},\vec{y})=\delta_{ij}\delta^3(\vec{x}-\vec{y}).
\end{align}.

Using the constraints matrix, we can construct the Dirac bracket of our model, which in turn will allow us to reduce our phase space. For the phase space functionals $F(t)$ and $G(t)$, we get
\begin{align}
    &\{F(t),G(t)\}_D= \{F(t),G(t)\}-\sum_{i,j=1}^2 \int d^3 \vec{z}  ~d^3 \vec{u} ~~\{F(t),~\phi_i(t,\vec{z})\}~\phi^{-1}_{ij}(\vec{z},\vec{u})~\{\phi_j (t,\vec{u}),~G(t)\}~.
\end{align}

Applying (\ref{constraintID}) it is easy to show that the Dirac bracket of the constraints with any phase space function is zero 
\begin{align}
     &\{F(t),\phi_k(t,\vec{x})\}_D= \{F(t),\phi_k(t,\vec{x})\}- \sum_{i,j=1}^2 \int d^3 \vec{z} ~d^3 \vec{u}~~ \{F(t), \phi_i(t,\vec{z})\}~\phi^{-1}_{ij}(\vec{z},\vec{u})~\phi_{jk}(\vec{u},\vec{x})=0,
\end{align}
this in turn allows us to take the constraints $\phi_1$ and $\phi_2$ as strong equalities and replace them in the total Hamiltonian
yielding our reduced Hamiltonian 
{\begin{align}
    \mathcal{H}_{red}=\int d^3x &\left[ \frac12 \pi_i^2+ \frac14 {f'_{ij}}^2
+\frac{1}{2M'}\left(\partial^i \pi_i +\pi^i V_i\right)^2
+a'_i V_j f'^{ij}+\frac12 M {a'_i}^2\right],
\end{align}
which indicates explicitly the presence of only the vector potentials and their conjugate momenta as phase spacce variables. It is also important to note that the Hamiltonian appears to be bound from below and as such the model is free of Ostrogradski ghosts, meaning that the model is classically well behaved.}

The total number of physical degrees of freedom of the system is then given by the master formula \cite{Henneaux-Teitelboum-1992} 
 \begin{align}
    2\!\times\! DOF &= ~ \text{Phase Space variables}\!-\! 2 \!\times\! \text{First Class Constraints}- \text{ Second Class Constraints}~.
\end{align}

The phase space is composed of $a'^\mu$ and $\pi^\mu$ and as a result we get 
\begin{align}
    2\times DOF = 8- 0-2 =6 \phantom{=} \Rightarrow DOF=3~,
\end{align}
which shows that there are three propagating degrees of freedom within the model. 

At this stage, one may now wonder whether we can still consider the 
model to be a gauge theory, as it is usually accepted that only first class constraints generate gauge transformations. There has been however recent results \cite{Pitts-2024} 
that show that even in the case of de Broglie-Proca electrodynamics, the second class constraints can generate gauge transformations. This reassures us with the notion that we are still dealing fundamentally with a gauge theory of electromagnetism {despite the manifest loss of gauge invariance}.\\
\subsection{Propagator}
{{  {The nature of the massive polarisation is however important to discuss, and to that extent we will try to analyse the propagator for the model. To strengthen our case we will consider the action \eqref{action 1} and put it in a suitable form 
\begin{align}
     \mathcal{L}'_0=& -\frac14f'^{\mu\nu}f'_{\mu\nu}-a'_\mu V_\nu f'^{\mu\nu}-\frac12 \left(a'^\nu a'_\nu V^\mu V_\mu- a'_\mu a'_\nu V^\mu V^\nu\right)&\nonumber\\
     =&\frac12 a'^\mu \Box \theta_{\mu\nu}a'^\nu+ \frac12 a'^\mu \left(V_\mu\partial_\nu-V_\nu\partial_\mu\right)a'^\nu-\frac12a'^\mu\left(V^2 \eta_{\mu\nu}-V_\mu V_\nu\right)a'^\nu &\nonumber\\
     =&\frac12a'^\mu\left[\left(\Box-V^2\right)\theta_{\mu\nu}-V^2 \omega_{\mu\nu}+V_\mu\partial_\nu-V_\nu\partial_\mu+ V_\mu V_\nu\right]a'^\nu&\nonumber\\
     =& \frac12 a'^\mu~ \Gamma_{\mu\nu}~a'^\nu~~,
\end{align}
where we replaced $\eta^{\mu\nu}=\theta^{\mu\nu}+\omega^{\mu\nu}$ with $\theta_{\mu\nu}$ and $\omega_{\mu\nu}$ representing respectively the transverse and longitudinal projector operators. We thus define
\begin{align}
    \Gamma_{\mu\nu}=\alpha~ \theta_{\mu\nu}+\beta~ \omega_{\mu\nu}+\gamma~V_\mu \partial_\nu + \delta~ V_\nu \partial_\mu +\epsilon~V_\mu V_\nu ~,
\end{align}
with 
\begin{align}
    \alpha= \Box-V^2 ,~~\beta= -V^2, ~~ \gamma=-\delta= 1, ~~ \epsilon=1&. \nonumber
\end{align}

The propagator being defined by \cite{baetaetal2004}
\begin{align}
    \langle0| ~T~[a'_\mu(x), a'_\nu(y)]~|0\rangle=i \left(\Gamma^{-1}\right)_{\mu\nu} \delta^4 (x-y),
\end{align}
we must now look for the inverse matrix $M^{-1}$.\\

After some rather tedious algebraic manipulations we can derive the following form 
\begin{align} \label{propagator}
    {\Gamma}^{-1}_{\mu\nu}= \frac{1}{\alpha}\theta_{\mu\nu}&+\frac{1}{\Delta}\left[\left(\alpha+\gamma V\cdot\partial+\epsilon V^2\right)\left(1+\frac{\delta}{\alpha}V\cdot\partial\right)+\left((\alpha-\beta)V\cdot\partial-\delta V^2\Box\right)\frac{\epsilon}{\alpha}\frac{V\cdot \partial}{\Box}\right]\omega_{\mu\nu}&\nonumber\\
    &+\frac{V_\mu \partial_\nu}{\Delta}\left[\left(-\gamma-\epsilon\frac{V\cdot\partial}{\Box}\right)\left(1+\frac{\delta}{\alpha}V\cdot\partial\right)+\left(\beta+\delta V\cdot \partial\right)\frac{\epsilon}{\alpha} \frac{V\cdot \partial}{\Box}\right]\nonumber\\
    &+\frac{V_\nu \partial_\mu}{\Delta}\left[-\left(\alpha+\gamma V\cdot \partial+\epsilon V^2\right)\frac{\delta}{\alpha}-\left(\alpha \frac{V\cdot\partial}{\Box}-\beta \frac{V\cdot\partial}{\Box}-\delta V^2 \right)\frac{\epsilon}{\alpha}\right] &\nonumber\\
    &+\frac{V_\mu V_\nu}{\Delta}\left[\left(\gamma \Box+ \epsilon V\cdot\partial\right)\frac{\delta}{\alpha}-\left(\beta+ \delta V\cdot\partial\right)\frac{\epsilon}{\alpha}\right],
\end{align}
with 
\begin{align} \label{pole2}
    \Delta \equiv  \alpha \left[\beta+\left(\gamma+\delta\right)V\cdot\partial+\epsilon \frac{\left(V\cdot\partial\right)^2}{\Box}\right]+\left(\frac{\beta \epsilon}{\Box}-\gamma\delta\right)\left(V^2 \Box-\left(V\cdot\partial\right)^2\right)~.
\end{align}

Replacing the values we set for the coefficients, we can see that $\Delta=0$ and such, we have a singularity on longitudinal modes. We must then introduce a gauge fixing term in the action of the form 
\begin{align}
    \mathcal{L}_{gf}=\frac{1}{2 \xi} \left(\partial_\mu a'^\mu\right)^2 =-\frac{1}{\xi}  a'^\mu\Box \omega_{\mu\nu} a'^\nu~,
\end{align}
this in turn changes the $\beta$ coefficient such as to have
\begin{align}
    \alpha=\Box-V^2 ,~~\beta= -\frac{\Box}{\xi}-V^2,~~\gamma=-\delta=\epsilon=1~,
\end{align}
which finally gives us for $\Delta$
\begin{align}
    \Delta= -\frac{1}{\xi}\left(\Box^2+\left(V\cdot\partial\right)^2\right)~.
\end{align}

With this in hand, we can state that there exist two poles for the propagator \eqref{propagator}} :
\begin{itemize}
    \item One on the transverse part $\theta_{\mu\nu}$ by setting $\alpha=0$ which gives
    \begin{align}
        \alpha=0 \Rightarrow k^2=-V^2= \vec{V}^2~.
    \end{align}
    \item And a second one on the longitudinal part $\omega_{\mu\nu}$ by solving $\Delta=0$ which reads
    \begin{align}
        k^2-\left(V\cdot k\right)= k^2+ \vec{V}\cdot\vec{k}=0~.
    \end{align}
\end{itemize}
\section{Discussion}

In this framework, we proved the existence of rest energy for photons which is usually considered as tied to mass. However, it appears to be associated to the transversal part of the propagator rather than longitudinal. An interesting follow-up to this work would be to develop the {  quantisation} 
of said model, and look into the {  Ward} identity and loop corrections, to see whether the longitudinal component could get shifted/corrected by higher order interactions.

However, we can state that classically it appears that the longitudinal component {  - while modulated - } does not get shifted in the rest frame by a non-dynamical background. Longitudinal here is thought in terms of the four-dimensional Minkowski space.
So, the mass induced by the variable non-dynamical background described by the constant vector $ V^\mu V_\mu$ is accommodated in the
transverse part of the photon propagator, corresponding then to the null helicity component of the spin-1 carried by the photon. While this might seem unusual, this behaviour is rather common for topologically induced masses, as topologically massive models are still manifestly gauge invariant. 

This in particular seems to indicate that while the model appears to be not gauge invariant, there must exist some transformation 
that would retrieve it. As such, the field redefinition described in \eqref{fieldredif} is only a formal symmetry breaking, and only the manifest gauge invariance has been lost, in the same manner that a Higgs mechanism would act on the vacuum state.\\

A very important point to underline concerns the mechanism behind the mass generation. In modern physics, mass is taken to be an emergent phenomenon in the most general case. Masses will emerge through interactions which will spontaneously break one of the symmetries (gauge symmetry, chiral symmetry etc.) of a model, which in turn generates Goldstone bosons which can be then "swallowed" by the particles to add a third polarisation (which can be longitudinal or transverse depending on the model). As such, it is generally accepted in physics that symmetry breaking can generate a mass.

In this model, the field redefinition at play appears to be generating a mass, and it could be argued
that such a redefinition, despite our arguments, is unjustified. 
Let us return to} 
the original form of the action before any modification. The coefficients coupling to the photon fields are not constants, which means that these dielectric coefficients are functions of space-time (in our case, space more specifically). This in turn has some consequences on the vacuum state of the system, which will develop a different refractive index and break its own symmetry. As such, while the original action seemed to obey two $U(1)$ invariances (photon and background), it is reasonable to believe that due to the non-dynamical nature of the background,  
a spontaneous symmetry breaking hidden at the quantum level would emerge. By choosing to absorb the order zero background contribution into the potential 4-vector, we are indirectly showing this effect by proceeding to a manifest symmetry breaking, which allows us to describe classically the otherwise subtle effects of the background field.  These mechanisms seem very close to those discussed in \cite{baetaetal2004} which is coherent as the constant nature of $V^\mu$ indicates a preferred direction in our system. \\

The construction of the constrained Hamiltonian formalism here is a first step towards the quantisation of the model. If pursued, we would have to consider the BRST operator of the model, ensure the unitarity at the quantum level and then couple the fields to fermions. Finding a fermionic representation might seem  simple, 
but we have to keep in mind that if the effects of the background on photons are {  not} trivial, they should also be non trivial on fermions. Once fully {  quantised}, and with the proper Feynman rules in hand, we could finally state 
if our mass is physical, or simply an effective polarisation that only holds at the classical level.

\section{Conclusions}
After having previously established the existence of a frequency shift in models where a photon-background split is executed, we study  in this work the conditions which  could induce an effective mass to a photon crossing an electromagnetic background in a generic non-linear model of electromagnetism. We notably show the emergence of a third propagating degree of freedom, as well as a modified constraint structure for the model, close to those of de Broglie-Proca and Stueckelberg. We show that the Hamiltonian of the model is bound from below, that the propagator possess well-behaved
poles and that our effective mass is definite positive.  The emergence of an effective mass in non-linear theories and in the Standard-Model Extension are strong motivations for investigating further the theoretical foundations of physics as well as the applications in astrophysics and particle physics.

\section{Acknowledgments}
Funding from ANR-DFG (Agence Nationale de la Recherche - Deutsche Forschungsgemeinschaft) for GMT (Generalised Maxwellian Theories ANR-22-CE92-0028-01) received together with Universit\"at Bremen is acknowledged. We would also like to thank C. L\"ammerzahl, V. Perlick and A. Shala for the discussions and corrections that they offered. 

\bibliography{references_spallicci_250703}

\end{document}